# Statistical Analysis of Submicron X-Ray Tomography Data on Polymer Imbibition into Arrays of Cylindrical Nanopores

Fernando Vazquez Luna,[#] Michael Gerstenberger,[#] Guido Dittrich,[§] Juliana Martins de Souza e Silva,[&,†] Patrick Huber,[§] Ralf Wehrspohn,[&] Martin Steinhart[*,#]

[#] Institut für Chemie neuer Materialien and CellNanOs, Universität Osnabrück, Barbarastr. 7, 49069 Osnabrück, Germany

[§] Hamburg University of Technology, Institute for Materials and X-Ray Physics, 21073 Hamburg, Germany and Deutsches Elektronen-Synchrotron DESY, Centre for X-Ray and Nano Science CXNS, 22603 Hamburg, Germany and Hamburg University, Centre for Hybrid Nanostructures CHyN, 22607 Hamburg, Germany

[&] Martin-Luther-Universität Halle-Wittenberg, Institut für Physik, Fachgruppe Mikrostrukturbasiertes Materialdesign, Heinrich-Damerow-Str. 4, 06120 Halle, Germany

## Abstract

Frozen transient imbibition states in arrays of straight cylindrical pores 400 nm in diameter were imaged by phase-contrast X-ray computed tomography with single-pore resolution. A semi-automatic algorithm yielding brightness profiles along all pores identified within the probed sample volume is described. Imbibition front positions are determined by descriptive statistics. A first approach involves the evaluation of frequency densities of single-pore imbibition lengths, a second one the evaluation of the statistical brightness dispersion within the probed volume as a function of the distance from the pore mouths. We plotted average imbibition front positions against systematically varied powers of the imbibition time and determined the optimal exponent of the imbibition time by considering the correlation coefficients of the corresponding linear fits. Thus, slight deviations from the proportionality of the average imbibition front position to the square root of the imbibition time predicted by the Lucas-Washburn theory were found. A meaningful preexponential factor in the power law relating imbibition front position and imbibition time may only be determined after ambiguities regarding the exponent of the imbibition time are resolved. The dispersion of peaks representing the imbibition front in frequency densities of single-pore imbibition lengths and in brightness dispersion profiles plotted against the pore depth is suggested as measure of the imbibition front width. Phase-contrast X-ray computed tomography allows the evaluation of a large number of infiltrated submicron pores taking advantage of phase-contrast imaging; artifacts related to sample damage by tomography requiring physical ablation of sample material are avoided.

## Introduction

Imbibition of porous scaffolds[1] involves the replacement of a receding fluid, such as air, by an invading fluid, such as a polymer. Attention has predominantly been focused on the imbibition of percolating pore systems, in which transient imbibition fronts consist of hydraulically coupled menisci.[2-5] The Laplace pressures across the menisci at the imbibition front depend on their curvatures. Local differences in the Laplace pressures caused by local pore diameter variations result in the occurrence of cooperative imbibition phenomena like imbibition front roughening,[6-9] viscous fingering[6, 8] and avalanche-like relaxations of the imbibition fronts.[8, 10-13] These cooperative phenomena superimpose on effects governing imbibition on the single-pore level. The imbibition kinetics of single cylindrical pores is phenomenologically described by a power law of the type:

$$L_s = v \cdot t_i^n \quad \text{(Equation 1)}$$

The imbibition length $L_s$ is the length of the segment of a cylindrical pore, which is filled by an invading liquid after an elapsed imbibition time $t_i$. The classical Lucas-Washburn theory[14-15] predicts a value of 0.5 for the exponent $n$ of $t_i$. If the exponent $n$ is set to 0.5, Equation 1 is suitable to describe processes that slow down with time – as it is commonly observed for the imbibition of fluids into pores. Therefore, the value 0.5 for $n$ is typically accepted without further questioning. Failure of the classical Lucas-Washburn law is commonly ascribed to the imbibition prefactor $v$, nature and magnitude of which have remained matter of debate.[16-17] For example, the effective pore diameters may be reduced by molecular layers immobilized on the pore walls, resulting in an increase in the effective viscosity.[18-21] Moreover, departures from macroscopic hydrodynamic behavior, and in particular from lamellar liquid flow, have been ascribed to thermal fluctuations, to van der Waals forces effective between invading fluid and pore walls, as well as to hydrodynamic slippage.[22] For polymers, the extent of slippage was found to depend on chain length, entanglement density and surface chemistry.[23-24]

Predictive understanding of imbibition is still premature. In 2010, Engel et al. argued that "...there exists no description of the filling behavior of liquids, especially of polymer melts, into nanopores in literature that is experimentally and theoretically consistent...".[16] In 2019, Singh et al. stated that "…there is a surprisingly sparse understanding of the processes occurring on the scale of individual pores and of how these processes determine the global invasion pattern".[25] Comprehensive predictive understanding of imbibition requires at first understanding of single-pore imbibition before even more complex cooperative imbibition phenomena related to percolation can be tackled. So far, either averaging methods, such as *in situ* small-angle X-ray scattering,[16] interferometric techniques[26] or dielectric spectroscopy,[27-29] have been employed to assess imbibition into ensembles of parallel, non-connected, cylindrical pores, or imbibition kinetics has been assessed by inspection of cross-sectional microscopic images with the naked eye.[30-31] The three-dimensional imaging of imbibition fronts of porous scaffolds containing submicron pores with single-pore resolution and the algorithmic evaluation of thus-obtained real-space images enabling quantitative comparison of experimental results and imbibition models have remained challenging. It would, in particular, be desirable to quantify the width of imbibition fronts, which is an important parameter for the validation of imbibition models.[9, 32]

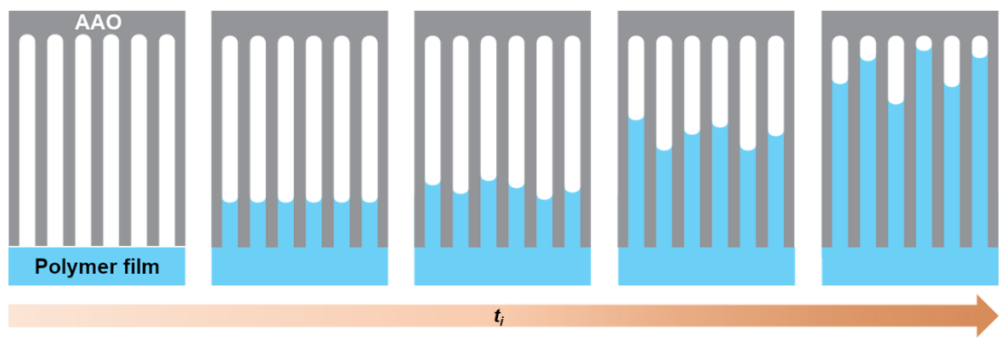

**Figure 1.** Schematic representation of an AAO template (gray) infiltrated with polymer (blue). With increasing imbibition time $t_i$ (from the left to the right), single-pore imbibition lengths $L_s$, the average position of the imbibition front $L_a$ as well as the width of the imbibition front increase.

Here we statistically analyze volumetric reconstructions of frozen transient stages of polymer melt imbibition into self-ordered nanoporous anodic aluminum oxide (AAO)[33-34] (Figure 1) with a nominal pore diameter of ~400 nm, a lattice constant of ~500 nm and a pore depth of ~100 μm. Self-ordered AAO is accessible by two-step anodization[33-34] and contains arrays of straight cylindrical pores with narrow pore diameter distribution.[35-36] The porous AAO layers with closed hemispherical pore bottoms were attached to underlying aluminum substrates. Imbibition of polymer melts into self-ordered AAO is of technological relevance since self-ordered AAO is routinely employed as template for the preparation of polymer nanorods by replication molding.[37-39] Polystyrene (PS) with a mass-average molecular weight $M_w$ = 239 kg/mol and a number-average molecular weight $M_n$ = 233 kg/mol was infiltrated into the AAO for infiltration times $t_i$ of 3, 10, 20, 30, 70 and 90 minutes at 200°C. At this temperature, each PS chain in a bulk PS melt is involved in ~13 entanglements.[40-41] Bent at al. reported for quiescent melts of PS with $M_w$ = 260000 g/mol, which is comparable to the $M_w$ value of the PS used here, at a temperature comparable to the infiltration temperature applied here a radius of gyration of 12 nm.[42] Under flow, the radius of gyration in the flow direction increased to 15 nm, while the radius of gyration normal to the flow direction slightly decreased to 11 nm. Therefore, it is reasonable to assume that the diameter of gyration is at least one order of magnitude smaller than the AAO pore diameter. It should be noted that trapped air did not influence the imbibition process since diffusive transport of small molecules, such as $N_2$ and $O_2$, is orders of magnitudes faster than the imbibition processes studied here.[43]

Three-dimensional reconstructions of the morphologies of the PS-infiltrated self-ordered AAO membranes were acquired by X-ray computed tomography[44-48] with submicron resolution in the phase-contrast mode. Since high-resolution X-ray computed tomography measurements last several hours, this method is not suitable for *in situ* imbibition monitoring. However, the highly entangled PS used here is characterized by a high glass transition temperature of ~100°C. Therefore, transient imbibition stages can conveniently be frozen by thermal quenching to room temperature, where the PS is vitrified. The quenched transient imbibition stages remain arrested throughout the X-ray computed tomography measurements. Consequently, the time scales of imbibition and the X-ray tomography measurements are decoupled.

To link volumetrically reconstructed sample morphologies to physical models, sample features of interest as well as parameters allowing their quantitative description need to be identified. So far, sample evaluation is frequently done by visual inspection – an approach associated with noise restricting its reliability.[49] We thus developed a semi-automated algorithmic three-stage image analysis procedure based on computer vision with a graphical user interface implemented in Python for the identification of the imbibition fronts, the determination of average imbibition lengths $L_a$ and the quantification of the imbibition front widths using open-source software packages. The results obtained in this way were compared with results obtained by manual evaluation of scanning electron microscopy (SEM) images of cross-sectional PS-infiltrated AAO specimens.

## Materials and Methods
### Materials
Monodisperse sec-butyl- and H-terminated PS ($M_w$ = 239 kg/mol; $M_n$ = 233 kg/mol; PDI = 1.03) was purchased from Polymer Standards Service (Mainz, Germany). Toluene (anhydrous; 99.8 %) was purchased from Sigma Aldrich. Silicon wafers were purchased from Siegert Wafer GmbH. Self-ordered AAO was prepared by a two-step anodization procedure reported by Masuda and coworkers.[34] Aluminum chips (diameter 40 mm, thickness 1 mm; purity > 99.99 %) were annealed for 3h at 500°C under argon and electropolished at room temperature with a mixture of 25 vol-% 60 wt-% $HClO_4$ and 75 vol-% ultrapure $C_2H_5OH$ pre-cooled to 6°C under moderate stirring (400 rpm) for 8 min at 4 A and 20 V. The anodizations were carried out at -1 °C in 1 wt-% phosphoric acid solution at 195 V. After the first anodization for 30 h, the formed alumina layer was etched with an aqueous mixture containing 1.8 wt-% $CrO_3$ and 7.1 wt-% of an 85 wt-% $H_3PO_4$ solution at 30 °C. The second anodization was carried out for about 40 h until the formed AAO layer had a thickness of 100 µm. The AAO pores were widened by isotropic etching with 10 wt-% phosphoric acid solution at 30 °C for 2h. As a result, AAO layers with a lattice constant of about 500 nm, a pore diameter of about 400 nm and a pore depth of about 100 µm attached to underlying 900 µm thick aluminum substrates were obtained.

### Infiltration of self-ordered AAO with PS
Approximately 100 µm thick (as determined by SEM) PS films were prepared by solution casting. 40 µL of a 10-wt% solution of PS in toluene was dropped on Si wafers previously cleaned three times by successive ultrasonication in acetone, ethanol and methanol. The films were dried overnight under ambient conditions and then at 80 °C for 18 h under a vacuum. Before infiltration, the PS films were removed from the Si wafers. The self-ordered AAO membranes attached to underlying alumina substrates were heated to 200 °C under argon for 10 min. Then, the unsupported PS films were placed on the AAO surfaces for the desired infiltration time under argon. To quench the infiltration, the samples were immersed in ice-cooled deionized water. The samples were then dried overnight at 40 °C.

**SEM cross-sectional images for method *SEM-mean***

Before imaging, the aluminum substrates of the PS-infiltrated AAO membranes were selectively etched with a solution containing 3.4 g $CuCl_2 \bullet 2H_2O$ and 100 mL concentrated HCl per 100 mL $H_2O$ at 0°C. The freestanding PS-infiltrated AAO membranes were then cleaved perpendicularly to the membrane surface. The samples were sputter-coated with a platinum-iridium layer three times at 20 mA for 15 seconds in a K575X Emitech sputter coater. SEM images were obtained using a Zeiss *Auriga* microscope applying an acceleration voltage of 3 kV and using secondary electron as well as in-lens detectors for image acquisition ensuring that the polymer bulk film, the infiltrated AAO pore segments and the empty AAO pores were visible in the images. Analysis of the images was performed with the software ImageJ.[50]

**X-ray computed tomography imaging**

Samples for X-ray computed tomography were prepared by laser-micromachining in a microPREP™ (3D-Micromac) device. Sub-millimeter cuts of the samples yielded the designed conical shapes with an average tip diameter of 50 µm atop a cube-shaped base (Figure S1). Each sample was glued on top of a metallic pin with a laser-cut fitting for this base, which was then placed in the sample holder. Imaging experiments were performed with a Carl Zeiss Xradia Ultra 810 device operating with a Cr X-ray source (5.4 keV) using phase-contrast imaging. A gold phase-ring with a thickness adjusted to produce a phase-shift of $3\pi/2$ of the non-diffracted X-ray beam was placed near the back focal plane of the zone plate. A total of 901 projections of 64 µm x 64 µm were obtained over 180° with an exposure time of 70 s and no detector binning, resulting in an isotropic voxel size of 64 $nm^3$. Sample drift was corrected using the adaptive motion compensation function of the operating software of the Xradia 810 Ultra device, which performs a quick scan with a limited number of projections after the main scan to correct large sample drifts. Three-dimensional reconstruction of the images was done through a filtered back-projection algorithm using the software XMReconstructor integrated into the Xradia 810 Ultra.

**Analysis of XRM data**

*A) Detailed description of stage 1: data import and preprocessing.* After volumetric reconstruction of the data acquired by phase-contrast X-ray computed tomography, a suitable sub-volume containing an array of intact AAO pores was manually cropped for further evaluation. The cropped volume was then rotated to align the AAO pore axes with the *z*-axis of a cartesian coordinate system employed as external reference framework using a standard implementation that employs bicubic interpolation for high-quality results.[51] As a result, stacks of *xy* slices oriented normal to the AAO pore axes were obtained. Then, the brightness values of the pixels constituting the *xy* slices in the selected sub-volume were normalized to one.

*B) Detailed description of stage 2: identification of the center coordinates of the AAO pores.* To determine the center coordinates of the AAO pores, for each evaluated AAO membrane a certain number of adjacent *xy* slices close to but ahead of the imbibition front cutting through empty AAO pore segments was selected (10 slices for $t_i$ = 3 min, 40 slices for $t_i$ = 70 min, 30 slices for all other $t_i$ values). To remove pixel noise, these adjacent *xy* slices were condensed into one *xy* slice by averaging the brightness values of pixels with the same *xy* positions along the *z* direction parallel to the AAO pore axes. Subsequently, adaptive histogram equalization was applied to enhance the local contrast and the result was smoothened using a Gaussian filter. The resulting averaged *xy* slice, in which the AAO pores appear as dark regions with good contrast to the bright AAO pore walls, was binarized applying a relative luminance of 30 % as threshold value.

The coordinates of the black foreground pixels in the binarized averaged *xy* slice represent a single 2D point cloud with multiple clusters representing the AAO pores. While most pores are separated, the binary image also contains clusters representing two or more contiguous AAO pores. Therefore, simple flood fill procedures for instance segmentation fail. To overcome this problem, an iterative custom-made variation of mean-shift clustering to identify the center coordinates of the AAO pores was carried out as follows. The brightness values of the binarized averaged *xy* slices were initialized as an input matrix for the first iteration by ascribing an integer value of 1 to black foreground pixels belonging to the AAO pores and an integer value of 0 to bright background pixels belonging to the AAO pore walls. We then defined a local neighborhood extending 11 x 11 pixels$^2$ around each black pixel. The coordinates of the neighborhood's center of mass were calculated as the frequency-weighted arithmetic means of the coordinates of the non-zero pixels inside the neighborhood. The initially considered black pixel in the center of its neighborhood was then shifted to the neighborhood's center of mass. This operation increased the integer value ascribed to the location of the center of mass by 1. This increase was done in a new matrix, which was initialized with a frequency value of zero for all pixels. Conducting this operation for all black pixels yielded a new output matrix, in which integer frequency values of 0, 1 or 2 were assigned to all pixel positions in the binarized averaged *xy* slice. The output matrix of the first iteration was then used as input matrix for the second iteration. In further iterations, the same steps as in the first iteration were repeated. Local neighborhoods were defined about each non-zero pixel; coordinates of the neighborhoods' centers of mass of were calculated as arithmetic means of the coordinates of the non-zero pixels inside the neighborhood weighted by the sums of the integer values of all pixels at a specific coordinate. The initially considered non-zero pixel in the center of its neighborhood was then shifted to the neighborhood's center of mass, so that the new integer value of the center of mass was increased by the integer value of the displaced center pixel. Iterations were carried out until the output matrices of successive iterations converged. As a result, the black pixels representing specific AAO pores collapsed into a single *xy* pixel position. The computational effort linearly scaled with the number of black foreground pixels in the binarized and averaged *xy* slice, because black foreground pixels belonging to specific AAO pores constituted small clusters with similar size and shape so that the convergence of the pixel clusters representing specific AAO pores required approximately the same time. As clusters may merge during an iteration, fewer pixels needed to be processed in subsequent iterations.

**Estimation of $L_s$ frequency densities from SEM data**

We calculated single-pore imbibition lengths $L_s$ based on AAO pore diameters obtained by the evaluation of a SEM image and the application of a Lucas-Washburn model modified by Yao et al.[21] to consider the presence of a dead layer of PS molecules immobilized on the AAO pore walls.

*A) Determination of AAO pore diameters by evaluation of a SEM image*

A scanning electron microscopy image of the surface of an AAO membrane with a nominal pore diameter of 400 nm and a pore depth of 1 µm was taken with a Zeiss Auriga device at 3 kV acceleration voltage and with 8 mm working distance. The SEM image acquired with an in-lens detector extended 2048 pixels x 1350 pixels and showed an image field of 87.52 µm x 57.69 µm (23.4 pixels/µm). The pore diameters were determined using the software ImageJ.[50] Brightness and contrast values were adjusted in such a way that the apparent pore areas did not change when small brightness and contrast changes were applied. The "Make binary" function was used to generate a binary image. Then, the "Analyze particles" function was used to determine the diameters of 19135 evaluated AAO pores. AAO pores touching the edges of the SEM image were not considered. We obtained a mean AAO pore diameter of 384 nm +/- 22 nm, corresponding to a pore radius $b_0$ of 192 nm (Figure S2).

*B) Estimation of the bulk melt viscosity $\eta_0$ of PS*

We used the bulk viscosity $\eta_0 = 3.5 \cdot 10^4 \, \text{Pa} \cdot s$ of PS with a molecular mass $M_w$ = 123 kg/mol determined for a temperature of 170 °C by Kim et al.[52] as reference value. We corrected the viscosity for the molecular mass using a power law suggested by Berry and Fox (Equation 2) assuming an exponent of 3.4:[53]

$$\eta_0 \propto M^{3.4} \rightarrow \left(\frac{M_1}{M_2}\right)^{3.4} = \frac{\eta_{0,1}}{\eta_{0,2}} \quad \text{(Equation 2)}$$

To correct for the temperature $T$, we used the Williams-Landel-Ferry (WLF) equation (Equation 3).[54] We first calculated the viscosity $\eta_{0,g}$ at the glass transition temperature $T_g$ of PS assuming $T_g$ = 273 K [55] using WLF parameters $C_1$ = 17.44 and $C_2$ = 51.6 K.[54] Then, the bulk viscosity $\eta_0 = 1.15 \cdot 10^4$ Pa·s of the PS used in this work at 200°C was calculated also using Equation 3.

$$\log\left(\frac{\eta_0(T)}{\eta_{0,g}}\right) = -\frac{C_1(T-T_g)}{C_2+T-T_g} \quad \text{(Equation 3)}$$

### C) Estimation of the thickness of the dead layer

Yao et al.[21] defined an effective radius $b_{eff}$ for cylindrical pores filled with polymers that is the difference of the nominal pore radius $b_0$ and the thickness $\Delta b$ of the dead zone formed by molecules immobilized on the pore walls (Equation 4):

$$b_{eff} = b_0 - \Delta b \quad \text{(Equation 4)}$$

The presence of a dead layer can than be considered by modyfying the Lucas-Washburn equation as follows (Equation 5):

$$L_s = \sqrt{\frac{b_{eff}^4 \gamma \cos\theta}{2\eta_0 b_0^3}} \sqrt{t_i} \quad \text{(Equation 5)}$$

$L_s$ is the single-pore imbibition length, $t_i$ the elapsed imbibition time, $\gamma$ the surface tension of PS and $\theta$ the equilibrium contact angle of PS on the AAO pore walls. To calculate $b_{eff}$ and $\Delta b$, we equated the first square root of Equation 5 with the imbibition prefactors $v$ determined by methods *CT-mean* and *CT-rms*, as described below (Equation 6):

$$v = \sqrt{\frac{b_{eff}^4 \gamma \cos\theta}{2\eta_0 b_0^3}} \quad \text{(Equation 6)}$$

We inserted the mean value $b_0$ = 192 nm obtained by the evaluation of a SEM image described above of into Equation 6. For $\gamma$ we assumed a value of 29 mJ/m² and for cos $\theta$ a value of 1.[55] In this way, we obtained for both methods *CT-mean* and *CT-rms* $\Delta b \approx$ 40 nm.

*D) Calculation of $L_a$ and the standard deviation σ of $L_s$*

The single-pore imbibition lengths $L_s$ were calculated by inserting the radii $b_0$ determined for single AAO pores by the evaluation of a SEM image, as described above, into Equation 5. We thus obtained $L_s$ frequency densities. We then calculated the average imbibtion lengths $L_a$ as mean values of the sets of $L_s$ values as well as the standard deviations σ of $L_s$ for the different imbibition times $t_i$.

## Results and discussion

### Analysis of X-ray computed tomography data

The SEM image displayed in Figure 2 shows the relevant features of a PS-infiltrated AAO membrane. At the bottom, a bulk PS film located on the AAO surface is seen as homogeneous grey area. In the center, AAO pore segments filled with PS are seen. The AAO surface between the bulk PS surface film and the part of the AAO membrane infiltrated with PS is marked by a green horizontal line. At the top, empty AAO pore segments are seen (the closed AAO pore bottoms are at the very top outside the image field). Figure 3 shows details of sections through the three-dimensional volumetric reconstruction of AAO infiltrated with PS for 3 minutes obtained by phase-contrast X-ray computed tomography. These details are reproduced with single-pixel resolution. Each pixel represents a data point in the volumetric reconstruction. The edge lengths of the pixels correspond to 64 nm. The detail of the section along the AAO pore axes displayed in Figure 3a shows the imbibition front. The orientation of the section is the same as that of the SEM image displayed in Figure 2. Furthermore, the section shows a similar sample region. Remarkable features are the dark dot-like areas approximately in the center, which mark the imbibition front within the AAO pores. Figure 3a demonstrates the advantages of the phase-contrast mode, in which phase boundaries, such as the boundary between PS and air at the imbibition front in the AAO pores, are highlighted. Figure 3b displays a detail of a *xy* slice, that is, a section normal to the AAO pore axes, which intersects empty AAO pore segments close to the imbibition front. In both sections shown in Figure 3 the AAO pores are clearly discernible.

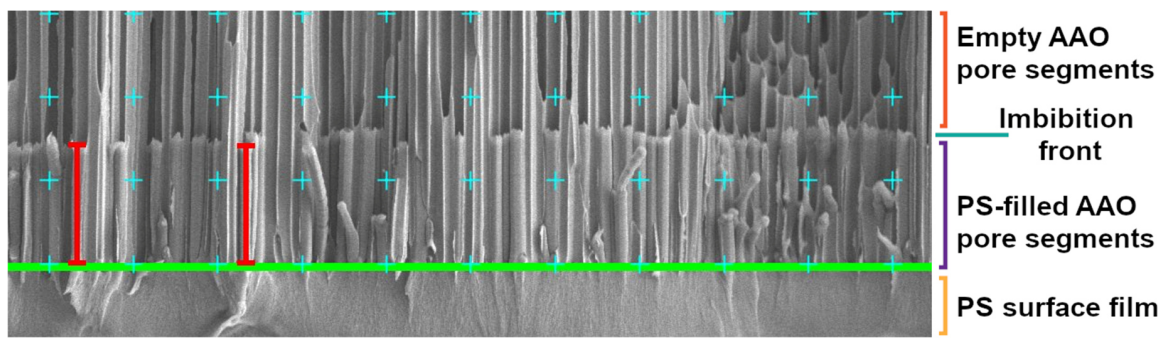

**Figure 2.** Scanning electron microscopy (SEM) image of an AAO membrane infiltrated with PS at 200°C for 3 minutes (detail of Supporting Figure S5a). At the bottom, a bulk PS film covering the AAO surface is seen. The green line marks the AAO surface. Above the green line in the center of the micrograph, PS-filled AAO pore segments and at the top empty AAO pore segments are seen. The closed bottoms of the AAO pores are at the very top outside the image. The grid marked by blue crosses helped align the AAO surface horizontally and the AAO pore axes vertically. The red lines indicate single-pore imbibition lengths $L_s$.

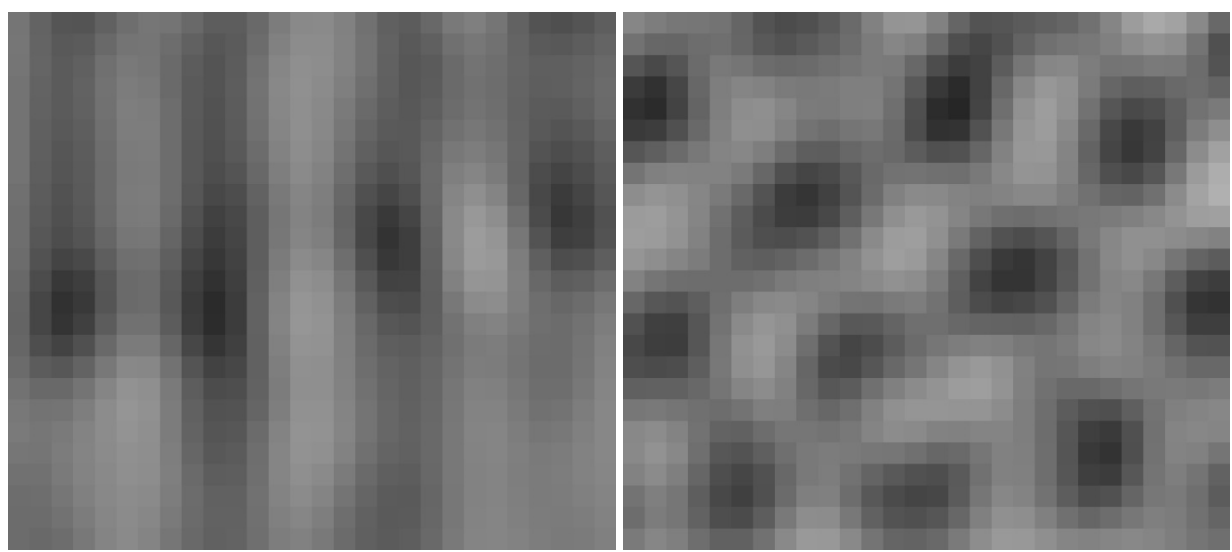

a)                                            b)

**Figure 3.** Sections with a width of 28 pixels and a height of 25 pixels through the volumetric reconstruction of AAO infiltrated for 3 min at 200°C with PS. Each pixel has edge lengths of 64 nm and corresponds to one data point. a) Section along the AAO pore axes showing the imbibition front. Empty pore segments are at the top, PS-filled AAO pore segments on the bottom. b) Slice normal to the AAO pore axes (*xy* slice) intersecting empty AAO pore segments close to the imbibition front. The empty AAO pores appear dark, the AAO pore walls bright.

It should be noted that the surfaces of the AAO pore walls may intersect pixels, which thus contain brightness contributions from the AAO pore walls and from the AAO pore volume. In general, a pixel can cover two or more objects, the boundary of which intersects the pixel. Therefore, we suggest to consider two times the pixel edge length, i.e., the minimum distance between a pixel center and the center of its second-nearest neighbor, as the resolution limit (here 128 nm). This resolution limit is, however, still sufficient to resolve single AAO pores. As obvious from a pixel intensity line profile along a part of a pixel row of Figure 3b shown in Figure S3, about 3 linearly arranged contiguous pixels with low pixel intensities represent the pores, which are separated by pore walls indicated by 1-2 pixels with high pixel intensities. In between, 1-2 pixels with intermediate pixel intensities with contributions from both the AAO pores and the AAO pore walls can be seen. The selected resolution allows the coverage of both the imbibition front and the AAO surface, as required to determine the imbibition lengths, and enables the evaluation of a large, four-digit number of AAO pores to ensure statistical reliability.

The challenge to be addressed in this work is the extraction of the relevant features from the volumetric reconstructions obtained by phase-contrast X-ray computed tomography, which include the position of the imbibition front relative to the surface of the AAO membrane (the average imbibition length $L_a$) and the width of the imbibition front. However, this problem cannot be solved by a simple segmentation procedure. Figure 4a shows a three-dimensional region of interest cropped from a volumetric reconstruction of the morphology of AAO, which was infiltrated with PS for 3 minutes at 200 °C. As in Figures 2, the PS surface film is located on the very bottom. The PS-filled AAO pore segments are seen in the lower part on top of the PS surface film and the part of the AAO membrane, in which the AAO pores are empty, on top. The contrast differences between empty and PS-filled pore segments as well as between PS-filled pore segments and AAO pore walls are minute. Moreover, brightness fluctuations occur not only between but also within the AAO pore walls as well as the PS-filled and the empty AAO pore segments. We thus devised a semi-automated algorithmic procedure to extract single-pore imbibition lengths $L_s$ from the volumetric reconstructions obtained by phase-contrast X-ray computed tomography, which can be divided into three stages (stages 1 and 2 are described in detail in section "Materials and Methods").

Stage 1 comprises data import and preprocessing involving the selection of a suitable sub-volume (Figure 4a) as well as alignment and normalization procedures. As a result, we obtained stacks of *xy* slices oriented normal to the AAO pore axes, which were in turn oriented parallel to the *z*-axis.

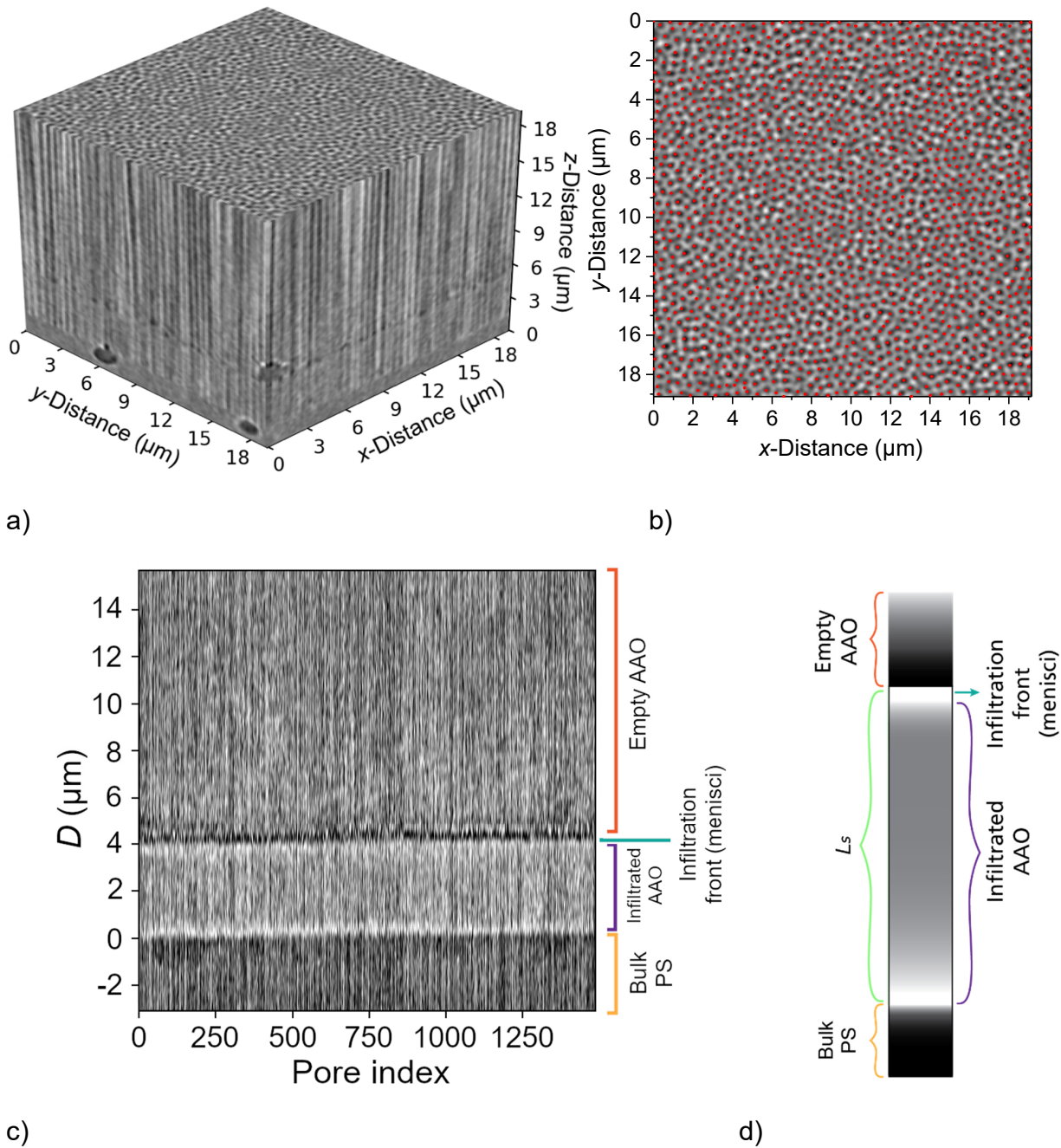

**Figure 4.** Phase-contrast X-ray computed tomography of an AAO membrane infiltrated with PS for 3 minutes. a) Cropped reconstructed volume of interest; b) View of a *xy* slice of the reconstructed volume of interest oriented perpendicularly to the AAO pore axes. The centers of the identified AAO pores are marked by red dots. c) Brightness tuple image, in which vertical pixel columns correspond to brightness tuples representing the brightness profiles along the identified AAO pores at the positions of the red dots in panel b). The distance to the AAO surface $D$ is plotted along the vertical axis, which is oriented parallel to the AAO pore axes. The horizontal axis represents an arbitrary index number ascribed to each considered AAO pore without physical meaning. d) Schematic representation of the brightness profile along a single AAO pore.

Stage 2 comprises the identification of the center coordinates of the AAO pores, which need to be determined with high precision because they are used to extract brightness profiles along the identified AAO pores. Thus, we generated a set of AAO pore center coordinates as follows. Several (here 10 – 40) adjacent *xy* slices intersecting empty AAO pore segments close to but ahead of the imbibition front were condensed into one "stage 2" *xy* slice. By this procedure, we improved the signal-to-noise ratio and the contrast between pore walls and pores. The condensed "stage 2" *xy* slice was binarized. Then, the center coordinates of the AAO pores were identified by nonparametric clustering applying an custom-made iterative variation of mean-shift clustering,[56] which enables fast computation with limited need of memory. In this way, the center coordinates of several thousand AAO pores were computed within seconds. Figures 4b and S4a show the positions of the AAO pores identified in a representative *xy* square area with an edge length of 19.134 µm. Figure S4b shows the Fourier spectrum obtained with Figure S4a as input image as well as the corresponding radial intensity profile rescaled into the real space. The radial intensity profile shows a characteristic peak representing the nearest-neighbor distance of the AAO pores. However, the obtained set of center coordinates does not contain all AAO pores present in the probed sample volume. Using the program ImageJ, we counted 1503 identified AAO pores, whereas the number of AAO pore expected for a perfect hexagonal lattice with a lattice constant of 500 nm would amount to 1691.

Stage 3 comprises the identification of the single-pore imbibition lengths. In each *xy* slice obtained in stage 1 the mean brightness values of 5 x 5 contiguous pixels surrounding the AAO pore centers identified in stage 2, which cover an area extending 320 x 320 nm$^2$, were calculated. The brightness values obtained for a specific AAO pore center from all *xy* slices were then arranged in a one-dimensional brightness tuple along the AAO pore axis (parallel to the *z* axis of the reference coordinate system). After linear detrending and normalization of the pixel brightness values, each tuple represents the brightness profile along the considered AAO pore. The brightness tuples obtained in this way were assembled into a brightness tuple image (Figure 4c), in which each vertical pixel column is a brightness tuple representing the brightness profile along one AAO pore (Figure 4d). Note that the sequence of the brightness tuples has no physical meaning.

The brightness tuple images differ from the volumetric reconstructions of the sample morphology directly accessible by phase-contrast X-ray computed tomography in that the contribution of the AAO pore walls is removed from the region of interest around the imbibition front. The removal of the contribution of the AAO pore walls is crucial for the identification of the imbibition front by phase contrast imaging because phase contrast only emerges at the menisci separating PS-filled and empty pore segments. Thus, the boundary between PS-filled and empty AAO pore segments is, as compared to the original volumetric reconstruction (Figures 3a and 4a), much more pronounced in the brightness tuple image seen in Figure 4c. The imbibition front appears in the brightness tuple images as a pronounced dark band oriented perpendicular to the brightness tuples and the AAO pore axes. The approximate position of the imbibition front was determined by visual inspection. Then, segments of the brightness tuples centering about the dark band were selected. The selected segments of the individual brightness tuples were linearly detrended. The *z* position of the darkest pixel within the detrended segments was determined as position of the imbibition front in the AAO pore represented by the considered brightness tuple.

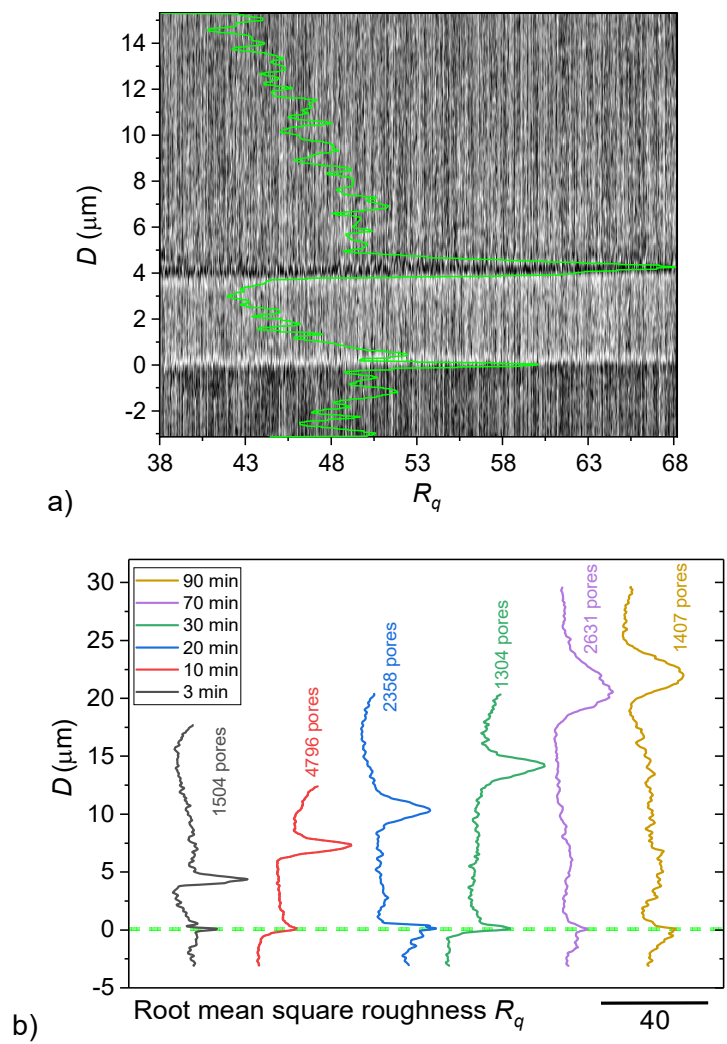

**Figure 5.** a) Brightness tuple image obtained from AAO infiltrated with PS for 3 minutes, on which the root mean square roughness $R_q$ of the brightness values in pixel rows oriented parallel to the horizontal axis plotted against the distance $D$ from the AAO surface (green) is superimposed. b) $R_q$ plotted against $D$ for different infiltration times $t_i$. The AAO surface is located at $D$ = 0 µm and indicated by a dashed green line. The indicated $R_q$ interval corresponds to a range of 40 out of 256 shades.

To calculate the single pore imbibition length $L_s$, the distance between the position of the imbibition front and the position of the AAO surface needs to be determined. In the brightness tuple images, the position of the AAO surface appeared as a bright band consisting of local brightness maxima in the brightness tuples. This outcome is, at first glance, unexpected because no interface separates the PS in the PS-filled AAO pore segments at the pore openings from the bulk PS film located on the AAO surface. However, the center coordinates of the AAO pores were determined from *xy* slices close to but ahead of the imbibition front. We assume that the brightness tuples trend the more away from the AAO pore axes the larger the distance to the imbibition front is. Considering that every pixel of the brightness tuples represents an area extending 320 x 320 nm$^2$ normal to the brightness tuple, it is straightforward to assume that the brightness tuples at the position of the AAO surface contain significant phase contrast contributions emerging from the interface between the AAO matrix and the bulk PS film located on the AAO surface. Thus, in the individual brightness tuples the position of the AAO surface is marked by local brightness maxima. The positions of these local brightness maxima were determined in the same way as the positions of the imbibition front. Therefore, trending of the brightness tuples away from the AAO pore axes is here rather an advantage than a drawback. As a result, for each brightness tuple, that is, for each AAO pore, $L_s$ could be determined as the distance between the brightness minimum representing the imbibition front and the brightness maximum representing the AAO surface.

**Average position of the imbibition front**

A straightforward approach to determine the average imbibition length $L_a$ (thereafter referred to *CT-mean*) is the calculation of the arithmetic mean value of the imbibition lengths $L_s$ of the individual AAO pores extracted from the corresponding brightness tuples. A second approach to determine $L_a$ (thereafter referred to *CT-rms*) involves the evaluation of horizontal brightness fluctuations in the brightness tuple images normal to the brightness tuples and the AAO pore axes, that is, parallel to the AAO surface. For this purpose, rows of pixels with the same *z* value oriented normal to the brightness tuples and the AAO pore axes (cf. Figure 4c) were evaluated. Each of these pixel rows contains exactly one pixel of each brightness tuple. Pixel rows behind the imbibition front at distances *D* from the AAO surface, where all AAO pores are filled with PS, consist of relatively bright pixels. Pixel rows ahead of the imbibition front at *D* values, where all AAO pores are empty, contain somewhat darker pixels. Because of the non-zero imbibition front width pixel rows at *D* values approximately corresponding to $L_a$ contain bright pixels belonging to AAO pore segments filled with PS ($L_s > D$), somewhat darker pixels belonging to empty AAO pore segments ($L_s < D$) and very dark pixels indicating the boundary between empty and filled AAO pore segments ($L_s \approx D$). The situation at the AAO surface is similar. Pixel rows belonging to the bulk PS film on the AAO surface ($D < 0$ µm) consist of relatively dark pixels, while the interface between AAO and bulk PS surface film is indicated by very bright pixels. However, since the position of the brightness maximum slightly varies between the individual brightness tuples, pixel rows at the approximate position of the AAO surface ($D \approx 0$ µm) contain dark pixels belonging to the bulk PS surface film, brighter pixels belonging to filled AAO pore segments and very bright pixels representing the interface between the AAO and bulk PS surface film. Therefore, particularly pronounced scattering of the pixel brightness values along the pixel rows normal to the brightness tuples representing the brightness profiles along the AAO pores indicates the position of the relevant interfaces, i.e., the imbibition front and the AAO surface (Figure 5a).

To identify the positions of the imbibition front and the AAO surface, we calculated the root mean square roughness $R_q$ of the pixel brightness values along rows of pixels normal to the brightness tuples in 8-bit brightness tuple images using the software Gwyddion[57] according to:

$$R_q = \sqrt{\frac{1}{N}\sum_{j=1}^{N}(h_j - \bar{h})^2} \qquad \text{(Equation 7)}$$

$N$ is the number of pixels per row, $h_j$ the gray value of pixel $j$, $\bar{h}$ the mean gray value of all pixels in the considered row and $j$ the index of summation. If $R_q$ is plotted against the distance $D$ from the AAO surface (Figure 5b), the average imbibition length $L_a$ corresponds to the distance between the $R_q$ extrema indicating the positions of the imbibition front and the AAO surface.

To validate the positions of the imbibition fronts extracted from X-ray computed tomography data, we evaluated for comparison cross-sectional 8-bit grayscale SEM images (Figure S5) in analogy to method *CT-mean*. We infiltrated three AAO membranes for each specific infiltration time $t_i$, under the same conditions. Then, we determined the imbibition lengths $L_s$, which correspond to the red vertical lines in Figure 2, for 10 pores per AAO membrane using the software ImageJ. Thus, overall 30 AAO pores of three different AAO membranes were evaluated for each $t_i$ value. This third approach is, thereafter, referred to *SEM-mean*.

**Quantification of the infiltration front width**

The imbibition fronts are indicated by peaks of image properties extracted from microscopic raw data as function of the distance $D$ to the AAO surface. The average imbibition front positions $L_a$ correspond to conspicuous points that can be determined unambiguously. However, the direct algorithmic determination of imbibition front widths would require the identification of the two imbibition front onset points. The onset of an imbibition front is indicated by a subtle change in the regional curvature of a raw curve exhibiting noise on local length scales. Any numerical $D$ values of discrete onset points will sensitively depend on how the indispensable curve fitting procedure is carried out. Moreover, the curvature change per distance interval $\Delta D$ in the onset region is close to zero over a larger $D$ range so that criteria for the rational positioning of discrete onset points can hardly be defined. Therefore, the direct determination of imbibition front widths is associated with significant uncertainties. Measures of the dispersion of the peaks indicating the imbibition front are more robust, albeit indirect, descriptors of the imbibition front width. For example, we calculated the standard deviations $\sigma$ of the frequency densities of the single-pore imbibition lengths $L_s$ obtained from phase-contrast X-ray computed tomography data (method *CT-mean*, Figure 6a) and by manual evaluation of SEM images as indicated in Figure 2 (method *SEM-mean*, Figure 6b). To ensure comparability between the different data sets, a uniform bin size $\Delta L_s$ = 0.5 µm was used for all histograms. For the third approach to quantify the widths of the imbibition fronts, we evaluated the root mean square roughness profiles $R_q(D)$ obtained by method *CT-rms*, which are displayed in Figure 5. As a measure of the imbibition front width, we determined the standard deviations $\sigma$ of Gaussian fits to the peaks representing the imbibition front (Figure 6c).

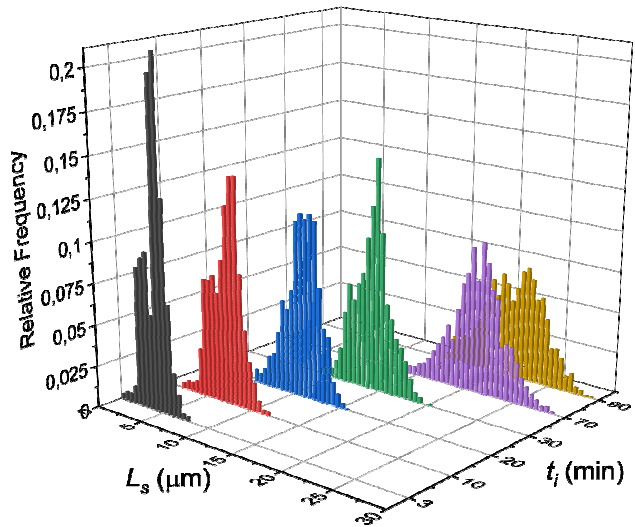

a)

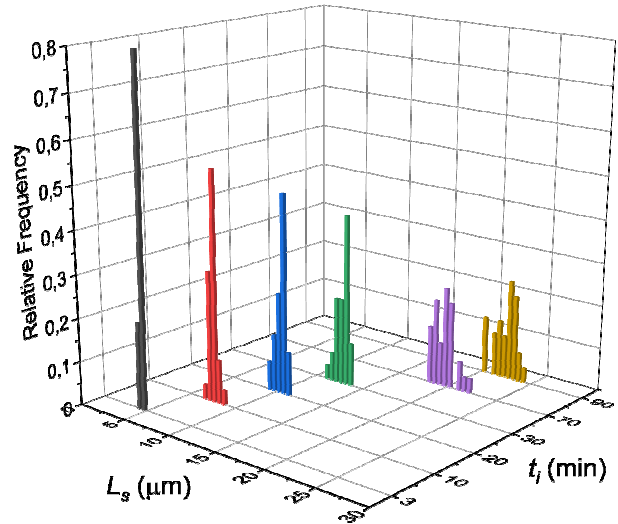

b)

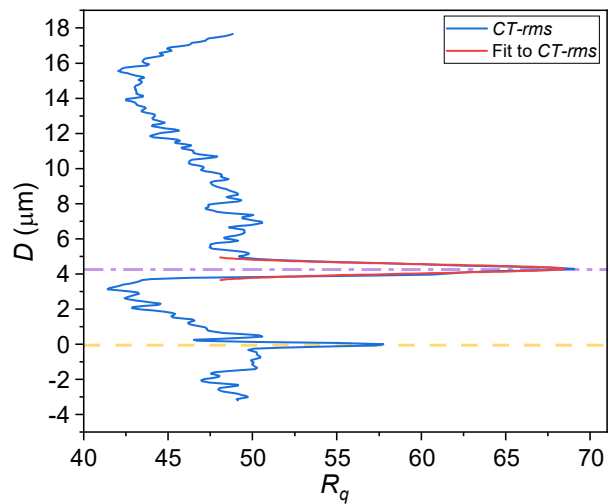

c)

**Figure 6.** Evaluation of the imbibition front width in PS-infiltrated AAO. a), b) Histograms displaying the relative frequency of the single-pore imbibition length $L_s$ (bin size $\Delta L_s$ = 0.5 µm) for different infiltration times $t_i$ obtained by methods a) *CT-mean* and b) *SEM-mean*. c) Gaussian fit (solid red line) to a $R_q(D)$ profile (root mean square roughness $R_q$ plotted against the distance $D$ from the AAO surface) obtained by method *CT-rms*. Here, the $R_q(D)$ profile for $t_i$ = 3 minutes is exemplarily shown. The dashed yellow line marks the position of the AAO surface and the purple dash-dot line the position of the imbibition front.

## Discussion

Real-space monitoring of the imbibition of PS into AAO by phase-contrast X-ray computed tomography with single pore resolution enables the evaluation of large numbers of AAO pores (here at least 1300) as well as the identification of menisci in the individual pores and directly yields frequency distributions of the single-pore imbibition lengths $L_s$. The number of AAO pores considered when cross-sectional SEM images are evaluated (method *SEM-mean*; cf. Table 1) is typically at least one order of magnitude lower. Moreover, the algorithmic semi-automated analysis of volumetric sample reconstructions obtained by phase-contrast X-ray computed tomography ensures unbiased identification of AAO pores in the probed sample volume. The results obtained by method *SEM-mean* involving the manual determination of single-pore imbibition lengths $L_s$ may be biased because of an implicit tendency to preferentially pick "good" AAO pores for evaluation.

The comparison of experimentally obtained measures for average imbibition lengths $L_a$ and imbibition front widths with predictions by theoretical models is far from being trivial. A first problem is the interpretation of experimental results. Even imbibition into straight cylindrical pores comes along with a complex phenomenology of transient imbibition states, including the presence of precursor films ahead of the imbibition front and voids in infiltrated pore segments behind the imbibition front.[58] Hence, there might be no "natural" absolute measures for the onsets, the positions and the widths of imbibition fronts. Moreover, quite different physical effects can be exploited for the imaging of imbibition fronts, and quite different statistical evaluation methods and measures can be used for the quantification of $L_a$ and the imbibition front widths. Depending on the applied methodology, different numerical values may be obtained. It is commonly assumed that imbibition lengths, be it the single-pore imbibition length $L_s$ or the average imbibition length $L_a$, scale with $t_i^{1/2}$, as predicted by the classical Lucas-Washburn theory. Indeed, at first glance all tested approaches seem to corroborate this notion. Plotting $L_a(t_i)$ against $t_i^{1/2}$ (Figure 7a and Table 1) seems to yield linear $L_a(t_i^{1/2})$ profiles and similar numerical values for the imbibition prefactor $\nu$ (cf. Equation 1), which corresponds to the slope of the linear fits to the different sets of $L_a(t_i)$ data points. We tested whether a value of 1/2 of the exponent *n* of $t_i$ indeed yields the best linear correlation between $L_a$ and $t_i^n$. A measure of the linearity of this correlation is the Pearson correlation coefficient $r$.[59] The closer $r$ to 1 is, the better is the correlation. We thus plotted $r$ as a function of *n* for the linear fits to the $L_a(t_i)$ data sets obtained by methods *CT-mean* and *CT-rms* (Figure 7b). For method *CT-rms*, the best linear correlation between $L_a$ and $t_i^n$ was obtained for $n \approx 0.45$. For method *CT-mean*, the best linear correlation between $L_a$ and $t_i^n$ was obtained for $n \approx 0.40$. The slopes $dL_a/dt_i^n$ of the linear fits to the sets of $L_a(t_i^n)$ data points obtained by methods *CT-mean* and *CT-rms* represent the imbibition prefactor $\nu$ of the power law describing the relation between imbibition length and imbibition time. As obvious from Figure 7c, the slopes $dL_a/dt_i^n(n)$ for exponents *n* ranging from 0.1 to 1.0 obtained by methods *CT-mean* and *CT-rms* nearly perfectly coincide. However, reducing the value of the exponent *n* from 0.5 to 0.4 resulted in a change in the apparent imbibition speed from ~2.4 µm/min$^n$ to ~4.1 µm/min$^n$ for both methods *CT-mean* and *CT-rms*.

The classical Lucas-Washburn theory does not account for the tendency of macromolecules, such as PS, to form interphases with reduced molecular mobility at substrate surfaces.[60] Indeed, the existence of so-called "dead zones" consisting of molecules of pore-invading polymers immobilized on the pore walls was reported previously.[31] Notably, the estimated thickness $\Delta b$ of the dead layers exceeded the radii of gyration of the invading polymer species several times. Yao et al. thus modified the classical Lucas-Washburn model considering the occurrence of dead layers by the introduction of a reduced effective pore radius $b_{eff}$ (equation 4) and an increased effective fluid viscosity $\eta_{eff}$.[21] Using an average AAO pore radius $b_0$ = 192 nm obtained by evaluation of a SEM image of the surface of an AAO membrane (Figure S2, for details see Materials and Methods), we equated the expression for the imbibition prefactor $v$ resulting from the modified imbibition model with the slopes $dL_a/dt_i^{1/2}$ of the linear fits to the sets of $L_a(t_i^{1/2})$ data points obtained by methods *CT-mean* and *CT-rms*. The results indicate the presence of a PS dead layer with a thickness $\Delta b = 40$ nm on the AAO pore walls. Using this value and the single pore diameter values determined by the above-mentioned evaluation of a SEM image, we applied the modified Lucas-Washburn model to calculate single-pore imbibition lengths $L_s$ for each AAO pore. The average values $L_a$ obtained from the $L_s$ frequency density for the different imbibition times $t_i$ (thereafter labelled "*Calc*") are in excellent agreement with the experimental values obtained by methods *CT-mean*, *CT-rms* and *SEM-mean*. This outcome is remarkable considering that the metrological precision of scanning electron microscopy is limited and that the values for the PS melt viscosity and the PS surface tension were approximated.

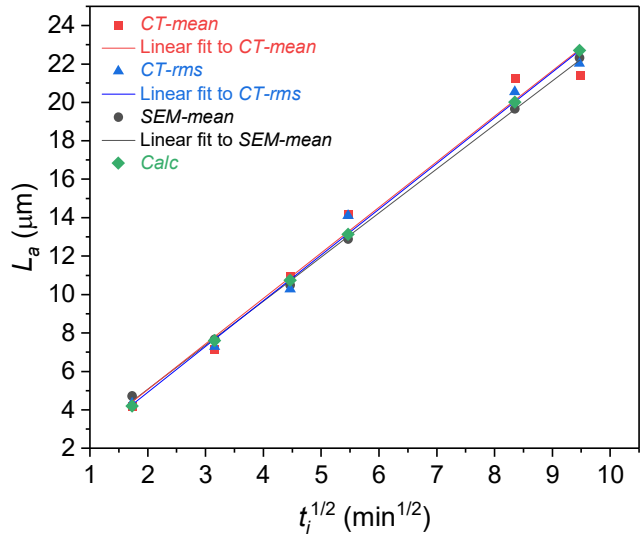

a)

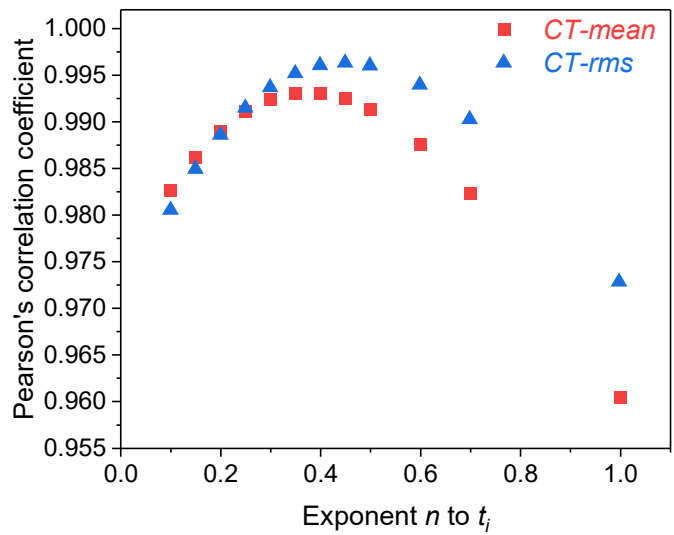

b)

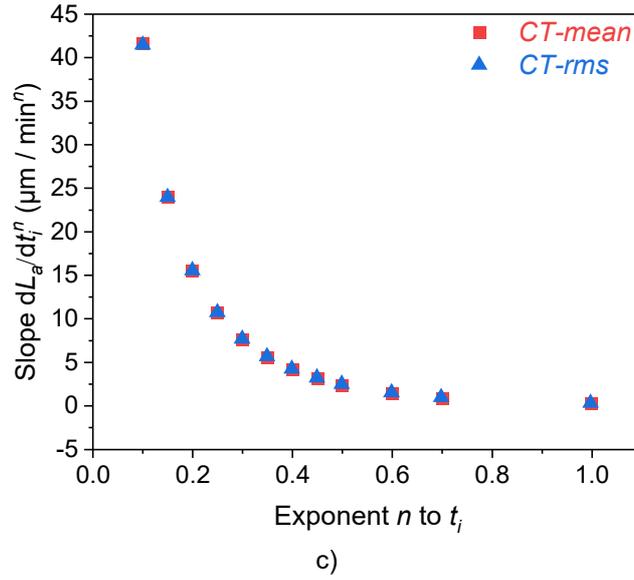

c)

**Figure 7.** Comparison of the average imbibition lengths $L_a(t_i)$. a) Sets of $L_a(t_i)$ data points obtained by methods *CT-mean* (red squares), *SEM-mean* (black circles) and *CT-rms* (blue up-triangles) as well as linear fits to the sets of data points plotted against the square root $t_i^{1/2}$ of the imbibition time $t_i$. For comparison, data points obtained by the evaluation of a SEM image of AAO and application of the modified Lucas-Washburn model devised by Yao et al.[21] are shown ("*Calc*", green diamonds). b) Pearson correlation coefficients of linear fits to sets of $L_a(t_i^n)$ data points obtained by methods *CT-mean* and *CT-rms* for exponents $n$ to $t_i$ ranging from 0.1 to 1.0. c) Slopes $dL_a/dt_i^n$ of linear fits to sets of $L_a(t_i^n)$ data points obtained by methods *CT-mean* and *CT-rms* for exponents $n$ ranging from 0.1 to 1.0 plotted against $n$.

The spatiotemporal evolution of the imbibition front width in the course of imbibition processes reveals morphological properties of the pore system being imbibed. Continuum-hydrodynamical concepts suggest that the single-pore imbibition prefactor $v$ (cf. Equation 1) depends on the balance of the flow-driving Laplace pressure and the pressure-drop caused by the viscous drag in the PS melt imbibing the AAO pores. As a result of the distinct dependencies of these contributions on the AAO pore diameters, the Lucas-Washburn kinetics for the advancement of the imbibition front is thus naturally transmitted into the broadening kinetics of the latter. The imbibition front width and measures thereof scale with a power function $t_i^n$ of the imbibition time $t_i$. The slope $d\sigma/dt_i^n$ of linear fits to sets of $\sigma(t_i)$ data points represents the scaling of the imbibition front width with the imbibition time $t_i$. The exponent $n$ reveals information on the shapes of the pores being imbibed. For a set of parallel pores with a random pore diameter distribution but uniform diameters along the individual pores the dispersion of the single-pore imbibition lengths $L_s$ scales with $t_i^{1/2}$.[9, 32] By contrast, for spontaneous imbibition into an ensemble of independent pores characterized by random diameter variations along the individual pores the imbibition front widening scales with $t_i^{1/4}$, as exactly derived by Grüner et al.[9]

**Table 1.** Numbers $N$ of the AAO pores considered per sample, Pearson correlation coefficients $r$ and slopes $v$ of the linear fits $L_a(t_i^{1/2})$ in Figure 7a as well as Pearson correlation coefficients $r$ and slopes $d\sigma/dt_i^{1/2}$ of the linear fits to the $\sigma(t_i^{1/2})$ data sets in Figure 8a.

| Method | N | $r$ value $L_a$ $(t_i^{1/2})$ | $v$ [µm/min$^{1/2}$] | $r$ value $\sigma(t_i^{1/2})$ | $d\sigma/dt_i^{1/2}$ [µm/min$^{1/2}$] |
|---|---|---|---|---|---|
| *CT-mean* | 1304-4796 | 0.991 | 2.372 | 0.994 | 0.205 |
| *CT-rms* | 1304-4796 | 0.996 | 2.380 | 0.929 | 0.146 |
| *SEM-mean* | 30 | 0.999 | 2.293 | 0.993 | 0.123 |

The direct algorithmic quantification of imbibition front widths is challenging. Methods *CT-mean*, *CT-rms* and *SEM-mean* yield raw curves of quantities accessible by evaluation of microscopic raw data, in which the positions of the AAO surface and the imbibition front are indicated by peaks. Statistical dispersion measures, such as standard deviations $\sigma$, allow the quantification of the widths of the peaks indicating the positions of the imbibition front and, indirectly, the quantification of the imbibition front width. However, the average imbibition length $L_a$ for the samples studied here amounts to a few µm up to several 10 µm, whereas the length scales relevant to the quantification of the imbibition front widths are one order of magnitude smaller and range from a few 100 nm up to a few µm. Therefore, the applied imaging method and the applied statistical data evaluation algorithm impact the quantification of the imbibition front widths more sensitively than the determination of $L_a$. Assuming that the pore diameter along the AAO pores is uniform, whereas AAO pore arrays are characterized by a certain pore diameter dispersion, we plotted $\sigma$ against $t_i^{1/2}$ and added linear fits to the different sets of $\sigma(t_i^{1/2})$ data points (Figure 8a and Table 1). For comparison, we calculated the standard deviations $\sigma$ of sets of $L_s$ values, which we calculated from the AAO pore diameters extracted from a SEM image using the modified Lucas-Washburn model reported by Yao et al.[21] The results obtained in this way are in reasonable agreement with the experimental results obtained by methods *CT-mean*, *SEM-mean* as well as *CT-rms* and corroborate the notion that a dead layer is present when PS is infiltrated into AAO. It would also be straightforward to calculate AAO pore diameters from the single-pore imbibition lengths $L_s$, if the classical Lucas-Washburn law was valid. However, the simple derivation of a pore diameter frequency density from an imbibition length frequency density is no longer possible if dead layers of PS molecules immobilized on the AAO pore walls are present. The equations resulting from the modified Lucas-Washburn model devised by Yao et al. can no longer be solved for the AAO pore diameter.

Plotting the Pearson correlation coefficient $r$ of the linear fits to the $\sigma(t_i)$ data sets obtained by methods *CT-mean* and *CT-rms* against the exponent $n$ (Figure 8b) yields $r(n)$ profiles that are vertically shifted. The $r(n)$ profiles reveal that the best linear correlation is obtained for $n = 0.60$ in the case of method *CT-mean* and for $n = 0.45$ in the case of method *CT-rms*, whereas the $r(n)$ values drop for smaller and larger $n$ values. The slopes $d\sigma/dt_i^n$ represent the scaling of the imbibition front widening with $t_i$. In the broader picture represented by Figure 8c, methods *CT-mean* and *CT-rms* yield slopes $d\sigma/dt_i^n$, which by and large coincide. For $n = 0.60$ in the case of method *CT-mean*, $d\sigma/dt_i^{0.60}$ amounts to 0.12 µm/min$^{0.60}$. For $n = 0.45$ in the case of method *CT-rms*, $d\sigma/dt_i^{0.45}$ amounts to 0.19 µm/min$^{0.45}$.

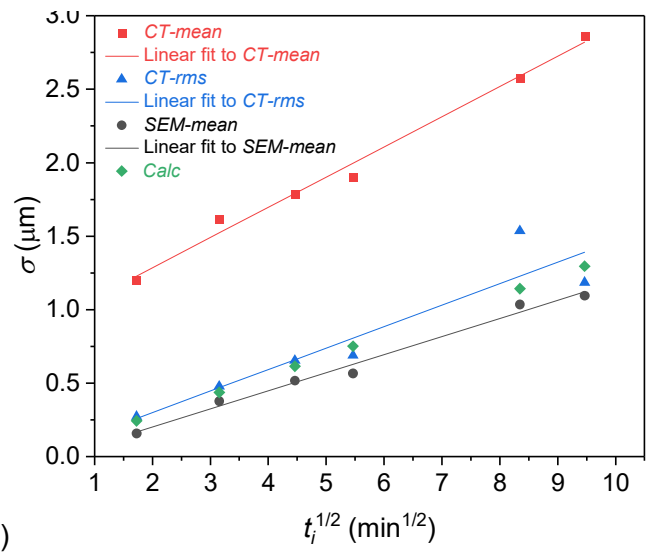

a)

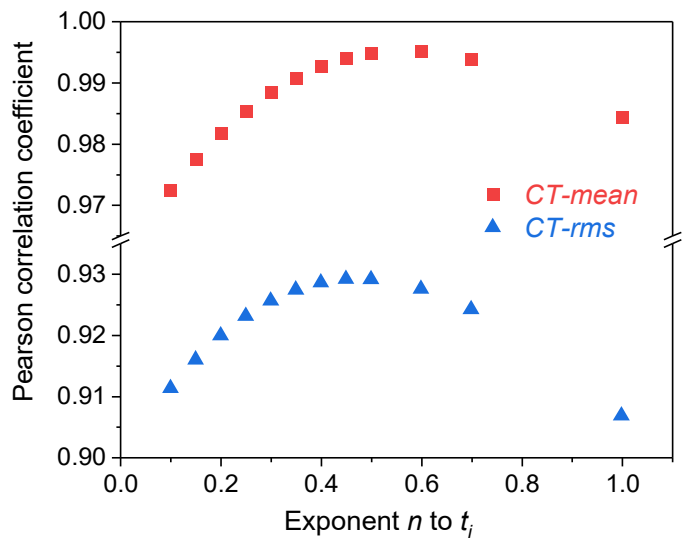

b)

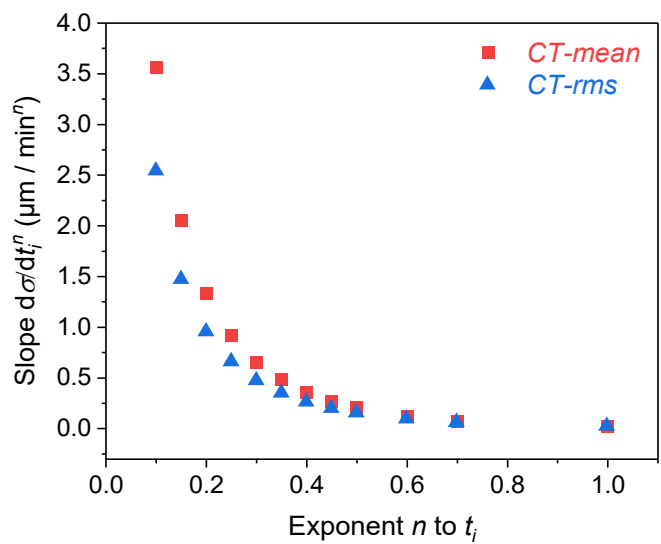

c)

**Figure 8.** Evaluation of the imbibition front widths as function of the imbibition time $t_i$. a) Standard deviation $\sigma$ as measure of the imbibition front width obtained by methods *CT-mean* (red squares), *SEM-mean* (black circles) and *CT-rms* (blue up-triangles) plotted against the square root $t_i^{1/2}$ of the imbibition time $t_i$. The solid lines are fits obtained by linear regression. For methods *CT-mean* and *SEM-mean* $\sigma(t_i)$ is the standard deviation of the single-pore imbibition lengths $L_s$ calculated from the histograms displayed in Figure 6a and b. For method *CT-rms* $\sigma(t_i)$ is the standard deviation of Gaussian fits to the peaks in the $R_q(D)$ profiles displayed in Figure 6c that represent the imbibition front. For comparison, standard deviations of $L_s$ frequency densities obtained by the evaluation of a SEM image of AAO and application of the modified Lucas-Washburn model devised by Yao et al.[21] are shown ("*Calc*", green diamonds). b) Pearson correlation coefficients of linear fits to sets of $\sigma(t_i^n)$ data points obtained by methods *CT-mean* (red squares) and *CT-rms* (blue up-triangles) for exponents $n$ ranging from 0.1 to 1.0. c) Slopes $d\sigma/dt_i^n$ plotted against $n$, the power of $t_i^n$.

## Conclusion

The understanding of liquid imbibition into porous scaffolds with submicron-sized pores is still premature – even for separated straight cylindrical pores, in which cooperative effects related to hydraulically coupled menisci are ruled out. Phase-contrast X-ray computed tomography resolving single pores with diameters in the submicron range may mitigate this knowledge deficit. As example, the infiltration of polystyrene into self-ordered AAO containing straight cylindrical pores 400 nm in diameter was studied. We developed a semi-automatic algorithm yielding brightness profiles along all identified AAO pores in the probed sample volume and used descriptive statistics to determine imbibition front positions and imbibition front widths. A first approach is based on the evaluation of histograms of single-pore imbibition lengths in the probed sample volume. For the second approach, the statistical brightness dispersion as a function of the distance to the AAO surface was evaluated. As compared to averaging methods to monitor imbibition without single-pore resolution, phase-contrast computed X-ray tomography allows direct microscopic real-space observation of the relevant sample features. As compared to FIB tomography, preparation artifacts are prevented, and phase-contrast imaging, which is particularly suitable for the detection of interfaces, can be deployed. It is commonly assumed that the imbibition length scales with the imbibition time to the power of 0.5, as predicted by the Lucas-Washburn theory. We determined the Pearson correlation coefficient of linear fits to sets of imbibition length/imbibition time data sets for different exponents and found the best linear correlation for exponents slightly deviating from 0.5. This outcome suggests that not only the imbibition prefactor in the Lucas-Washburn equation but also the exponent of the imbibition time should be scrutinized.

The scaling of imbibition front widths with the elapsed imbibition time represents the dispersion of the AAO pore diameters and yields information on the pore morphology. However, it is demanding to unambiguously identify the two imbibition front onset points for the direct determination of imbibition front widths. Instead, dispersion measures of peaks indicating the presence of imbibition fronts in plots of suitable image properties against the distance from the AAO surface were used to quantify imbibition front widths. Since the relevant length scales are one order of magnitude smaller than the length scales relevant to the determination of the mean imbibition front position, numerical values of measures of the imbibition front width stronger depend on the applied imaging method and the statistical evaluation methodology. Hence, a naïve comparison of numerical values obtained by different experimental methodologies is misleading. To get comparable results, clearly defined, reproducible algorithms comprising the measurement method and statistical data evaluation need to be applied. This implies that attention needs to be directed to the compatibility of experimental measures of imbibition front widths and theoretical imbibition models. The results of methods *CT-mean* and *CT-rms* obtained with AAO as well-characterized pore model validate theoretical predictions by Grüner et al.,[9, 32] according to which the imbibition front width scales with $t_i^{1/2}$ if pore models containing cylindrical pores uniform in diameter are used. Also, our results corroborate the notion that a dead layer of macromolecules immobilized on the AAO pore walls influence imbibition dynamics. Future works may address imbibition into more complex porous scaffolds containing spongy continuous pore systems with a tortuosity larger than 1.

ASSOCIATED CONTENT
**Supporting Information.** Figure S1: Images of a sample specimen for X-ray computed tomography. Figure S2: Histogram of AAO pore diameters obtained by evaluation of a SEM image. Figure S3: Pixel intensity line profile taken from an *xy* slice of a volumetric reconstruction of PS-infiltrated AAO. Figure S4: Positions of AAO pore center coordinates identified by nonparametric clustering and corresponding Fourier spectrum. Figure S5: SEM cross-sectional images of AAO membranes infiltrated with PS. Supporting Material: Software used to analyze computed X-ray tomography results.

AUTHOR INFORMATION
**Corresponding Author**
* Martin Steinhart, Institut für Chemie neuer Materialien und CellNanOs, Universität Osnabrück, Barbarastr. 7, 49076 Osnabrück, Germany; e-mail: martin.steinhart@uos.de

**Present Addresses**
† Fraunhofer Institute for Microstructure of Materials and Systems IMWS, Walter-Hülse-Straße 1, 06120 Halle (Saale), Germany

**Author Contributions**
The manuscript was written through contributions of all authors. All authors have given approval to the final version of the manuscript.

ACKNOWLEDGMENT
The authors thank the German Research Foundation for funding (PAK 949 "Nanostructured Glasses and Ceramics", project number 383411810, HU 850/9-1, STE 1127/19-1, WE 4051/22-1 and "High-throughput, Chemical X-ray Microstructure Screening Center for Functional Glasses and Glass Ceramics", project number 316987262, WE 4051/21-1) as well as Dr. Cristine S. de Oliveira, Claudia Hess and Christine Schulz-Kölbel for technical support.